\documentclass[12pt]{iopart} 
\usepackage{graphicx}
\usepackage[table]{xcolor}
\usepackage{rotating}
\usepackage{multirow,colortbl}
\usepackage{url}
\usepackage{lineno}

\newcommand{\Cellhi}{\cellcolor{gray}}
\newcommand{\Cellmed}{\cellcolor{lightgray}}

\makeatletter
\long\def\@makefntext#1{\parindent 1em\noindent 
 \makebox[1em][l]{\footnotesize\rm$\m@th{\arabic{footnote}}$}%
 \footnotesize\rm #1}
\def\@makefnmark{\hbox{$^{\arabic{footnote}}\m@th$}}
\def\@thefnmark{\arabic{footnote}}
\makeatother
\begin{document}

\title{The three most common needs for training on measurement uncertainty}

\author{Katy Klauenberg$^1$, Peter Harris$^{2}$, Philipp M\"ohrke$^3$ and Francesca Pennecchi$^4$}
\address{$^1$Physikalisch-Technische Bundesanstalt, Abbestr. 2-12, 10587 Berlin, Germany}
\address{$^{2}$National Physical Laboratory, Hampton Road, Teddington TW11 0LW, UK}
\address{$^{3}$Universit\"at Konstanz, Fachbereich Physik, 78464 Konstanz, Germany}
\address{$^4$Istituto Nazionale di Ricerca Metrologica – INRiM, Strada delle Cacce 91, 10135, Torino (TO), Italy}
\ead{Katy.Klauenberg@Ptb.de}

\begin{abstract} 

 Measurement uncertainty is key to assessing, stating and improving the reliability
of measurements. An understanding of measurement uncertainty is the basis for confidence in measurements and is required by many communities; among others in national metrology institutes, accreditation bodies, calibration and testing laboratories, as well as in legal metrology, at universities and in different metrology fields. An important cornerstone to convey an understanding of measurement uncertainty is to provide training. 

 This article identifies the status and the needs for training on measurement uncertainty in each of the above communities as well as among those teaching uncertainty. It is the first study to do so across many different disciplines, and it merges many different sources of information with a focus on Europe. As a result, awareness on the training needs of different communities is raised and teachers of uncertainty are supported in addressing their audiences' needs, in improving their uncertainty-specific pedagogical knowledge and by suggestions for training materials and tools. 

 The three needs that are most commonly encountered in the communities requiring an understanding of measurement uncertainty, are 
1) to address a general lack of training on measurement uncertainty, 
2) to gain a better overview of existing training on measurement uncertainty in several communities, and 
3) to deliver more training on specific technical topics including use of a Monte Carlo method for propagating probability distributions and treating multivariate measurands and measurement models.

These needs will serve to guide future developments in uncertainty training and will, ultimately, contribute to increasing the understanding of uncertainty.
\end{abstract} 
\noindent{\emph{Keywords:} measurement uncertainty, education, training\\}
\maketitle

\section{Introduction to Measurement Uncertainty}\label{sec:intro} 
	Measurement uncertainty is key to assessing, stating and improving the reliability of measurements. It is the basis for ensuring that measurement results are metrologically traceable to the SI \cite{JCGM200,MRAP11} and thus critical to metrology for the European Grand Challenges in health, environment and energy \cite{EurametSRA,Brown2021}, as well as for the quality infrastructure \cite{QI4SD}, industry \cite{EurametIndustry, EMUEComp}, trade, regulations \cite{MRA,JointDeclar}, etc. 
	
	Guidance on the evaluation of measurement uncertainty is available in the suite of documents of the GUM (the Guide to the expression of Uncertainty in Measurement \cite{JCGM100,JCGM101,JCGM102,JCGM106,GUM1,GUM6}). This guidance is published in the name of the international organizations BIPM, IEC, IFCC, ILAC, ISO, \mbox{IUPAC}, \mbox{IUPAP} and OIML, is applicable to a broad spectrum of measurement problems (see section~\ref{subs:EMNs} and e.g.\ \cite[sec.\ 1.1]{JCGM100} \cite[sec.\ 1]{GUM1}) and rests on sound principles of probability and statistics \cite{Bich2016}. Furthermore, the GUM is adopted by national metrology institutes (see section~\ref{subs:MU}), it is widely accepted \cite[sec.\ 2.1.1]{Bich2016} and required in accreditation \cite{ILACP10,JointDeclar}, for the statement of uncertainties on calibration certificates \cite{ILACP14,EA-4/02} and of calibration and measurement capabilities \cite{MRAP11}. The GUM is recommended when assessing conformance to tolerances \cite{ISO17025} and is the basis for guidelines, standards and policy documents in many application areas, e.g. \cite{ILACP14,EA-4/02,NCSLIRP12,UKAS3003,Eurachem}.

 An understanding of measurement uncertainty from the highest scientific level all the way down to the shop floor \cite[sec.\ 1.1]{JCGM100} is the basis for confidence in measurements. The need for a better understanding was expressed repeatedly, e.g.\ in a survey \cite{JCGMSurvey} and during a research project \cite{EMUESumm}. The communities with such a need are multifarious, because uncertainty accompanies diverse measurements. 
The remainder of this section lists the key communities having requirements for understanding and evaluating measurement uncertainty (section~\ref{subs:Comm}), indicates different ways to provide that understanding (section~\ref{subs:MU}), and describes the overall aims of the present study to support that provision from the perspective of training (section~\ref{subs:paper}).

\subsection{Communities Requiring an Understanding of Uncertainty}\label{subs:Comm} 
 	To structure the discussion of communities who require to understand measurement uncertainty, the article follows the metrological traceability hierarchy and focuses on key organizations of (national) metrology systems. These are 1) national metrology institutes, designated institutes and other signatories of the CIPM MRA \cite{MRA}, 2) accreditation bodies and technical assessors, 3) calibration laboratories, 4) testing laboratories, 5) legal metrology authorities and their organisations. 
		In addition, 6) students, lecturers and researchers at universities, 7) different metrology communities as represented by EURAMET's European Metrology Networks (EMNs) and Technical Committees (TCs), and finally 8) the teachers of measurement uncertainty themselves
 	require an understanding of measurement uncertainty. 

  Other communities may have related or different requirements for understanding uncertainty. For example, schoolteachers and students \cite{Munier2013,Premier2018}, standardization bodies, industrial companies (e.g.\ manufacturers interested in stating product specifications or confirming test equipment suitability), needs emerging in the digital transformation \cite{Eich21} and other fields new to metrology, and finally society in general have requirements to understand uncertainty at some level. However, these requirements will not be explicitly covered in this article. 

 \subsection{Support for Understanding Uncertainty}\label{subs:MU} 
	
	Besides guidance documents such as the GUM, there are other possible ways to increase or support the understanding of the concepts, evaluation, usage and reporting of measurement uncertainty. These possibilities are research, examples and software, as well as training.
	
	Methods for evaluating uncertainty are an active area of research, and those covered as well as those not covered by the GUM documents are continuously investigated and developed (c.f.\ \cite{MathmetSRA} and section~\ref{subs:NeedEmerg} ). 
	Also research on applications \cite{MathmetSRA} as well as on didactics (see section~\ref{subs:Teach}) of uncertainty increase the understanding of it at the highest level. Altogether, more than 3\,500 publications included the keyword `measurement uncertainty' in their title in the past decade (according to Google Scholar search).
	
	Worked out examples can illustrate the evaluation and reporting of measurement uncertainty and hereby aid practitioners' understanding or demonstrate good practice. Many examples can be found throughout the GUM documents \cite{JCGM100,JCGM101,JCGM102,GUM6} and in a future dedicated document \cite{JCGMNews}, as well as in many derived guidelines, standards, trainings \cite{ExSurvey} and in a compendium \cite{EMUEComp}. However, the principles underlying these examples need to be understood to properly adapt them to measurement models or conditions other than those given.
	
	A multitude of diverse software \cite{WikiMUSoftware} exists to compute uncertainty according to the GUM \cite{SurveySoftware}. This software ranges from user-friendly web applications or GUIs to comprehensive collections of libraries, it implements methods of one or several of the GUM documents, represents a broad scope software or a very tailored one. Some software provides evidence of its validation (c.f.\ \cite{Penn2023}). Generally, dedicated software may simplify or even partially automate the evaluation of uncertainty. Its users, however, still require an understanding of the principles.
	
An important cornerstone to increase the understanding of measurement uncertainty is to provide training courses to relevant communities. The next section will describe how this article will contribute to improve training and thus the understanding of uncertainty. 

 \subsection{Training on Uncertainty -- Content and Aim of This Article}\label{subs:paper}
	
 This work will identify training needs in communities that require an understanding of uncertainty. For this purpose, section~\ref{sec:stateofart} will describe the requirements for  understanding uncertainty and the current provision of training for each identified community. It also includes an overview on available resources for uncertainty training. Section~\ref{sec:needs} will then identify gaps in current training on uncertainty and list training needs for each community as well as overall needs that may emerge due to new developments in metrology. Finally, section~\ref{sec:Concl} will summarize and identify those needs most common to all communities and give recommendations for future developments in uncertainty training.
	
	The aim of this work is to raise awareness for the trainees from each of the communities that require an understanding of uncertainty, to highlight their training needs and identify common ones across the communities. These aspects will support the teachers of uncertainty to better know their audiences and to better address their needs. The listed resources for uncertainty training will also support teachers. The combination of increased awareness, identified needs, listed resources as well as recommendations may serve to guide future developments by providers of uncertainty training. 
	
	Ultimately, this work will contribute to increasing the understanding of uncertainty.
	
 Information will focus on measurement uncertainty training offered in Europe. It is based on surveys, interviews and a workshop (see \ref{AppSurvey} to \ref{AppWelm}) that were conducted in the scope of the project Measurement Uncertainty (MU) Training (an activity of the European Metrology Network  for Mathematics and Statistics, see \cite{MUTrain}). Information is also based on letters from and meetings with stakeholders of this project as well as on protocol statements from project partners (see \ref{AppSL}, stakeholder letters are labelled [SLn] and protocol statements [PS]). In addition, information from several European Metrology Networks will be used (see \ref{AppEMN}) to describe high-level needs in different metrology fields. Also research will give insights into uncertainty training.

  This article is the first to describe training on measurement uncertainty across many different disciplines. The study merges many different sources of information, but the authors do not claim complete coverage of all needs for all audiences. Inevitably, the study is limited by the sources of information used to describe the state of the art and the needs in uncertainty training. Although the completeness and quality of each of the sources varies, 
	 the authors think that the set as a whole provides the reader with a good overview. Those sources that are available publicly are referenced. Those sources that cannot be published are described in \ref{AppSource} to transparently link the information to conclusions drawn from it. 	
	
	The authors also hope to stimulate discussion in those communities not covered here. 

\section{State of the Art of Training on Measurement Uncertainty in Europe}\label{sec:stateofart} 
Training is usually based on the GUM or derived documents, often illustrated by examples and supported by software (c.f.\ \cite{Penn2023,CourseSurvey,ExSurvey}). 

This section will provide a more detailed overview on the state of the art in uncertainty training, considering separately each of the 8 communities listed in section~\ref{subs:Comm}. Each subsection starts with the community's requirement for evaluating and understanding uncertainty, which may itself be very heterogeneous.

 \subsection{NMIs, DIs and other Signatories of the CIPM MRA}\label{subs:NMI}
	
	The 252 institutes participating in the CIPM MRA, i.e.\ 97 national metrology institutes (NMIs), 151 designated institutes (DIs) and 4 international organizations \cite{MRAP13,MRAPartic}, mutually recognize their calibration and measurement capabilities, which are expressed in terms of measurement uncertainty \cite[paragr.\ T2, T7]{MRA}, \cite{MRAP11} and should comply with the GUM (see \cite[Note 4]{MRAG13}, also \cite[Note 3 in 7.6]{ISO17025}). These institutes therefore require a deep understanding of uncertainty for research, for participation in key comparisons and for disseminating traceability. Some larger NMIs employ teams of mathematicians and statisticians who expand, apply and teach methods to evaluate uncertainty (c.f.\ \cite[authors]{MathmetSRA}). But also the management and possibly even the administration of institutes participating in the CIPM MRA require some understanding of uncertainty.

  In Europe, at least the national metrology institutes in Spain (CEM), Poland (GUM),	Italy (INRiM), Portugal (IPQ), France (LNE), Switzerland (METAS), the UK (NPL), Ireland (NSAI), Germany (PTB), Sweden (RISE) and Belgium (SMD), as well as the European Commission's Joint Research Centre (JRC) each teach courses on or including measurement uncertainty. A recent survey (see \ref{AppSurvey} and \cite{CourseSurvey, Penn2023}) provides an overview on these 34 courses, 
	which comprise 15 courses aimed at audiences that include NMI staff. Out of these 15 courses\footnote{\label{note1}Courses are double counted when they fulfil several of the listed characteristics.}, 9 are dedicated to measurement uncertainty, 5 to metrology and one to calibration. Table~\ref{TabSurv} gives the numbers of these uncertainty courses that teach different technical content.
	Of these 15 courses, 13 use one or more examples and 10 use software in class. 7 courses are available online with most of these in prerecorded form. 12 of the courses provide some sort of certification.
		
	\begin{table*}
		\caption{\label{TabSurv}Numbers of courses teaching different technical aspects of measurement uncertainty for audiences including staff from NMIs, calibration or testing laboratories, in legal metrology and universities (from~\cite{CourseSurvey}). For reference, the equivalent question (Q) number for the questionnaire in table~\ref{TabAccredTopics} is given.}
			\begin{indented}\item[]
			\begin{tabular}{@{}lr|rrrr@{}}
			 & \mbox{\hspace{0em}equiv.\ topic}& \multicolumn{4}{c}{audience}\\
			 technical content of uncertainty course& in table~\ref{TabAccredTopics} & NMI & lab & legal & uni \\\hline
			 mathematical tools reviewed / as prerequisite & Q1 & 7/7 & 9/7 & 0/2 & 6/6\\
			 (some) probability concepts & Q2 & 15 & 22 & 6 & 12\\
			 basic metrological concepts & Q3 & 15 & 23 & 6 & 15\\
			 standard uncertainties for input quantities & Q4 & 15 & 24 & 6 & 16\\
			 law of propagation of uncertainty (LPU) & \mbox{\hspace{-2em}Q5 (maybe Q6)$\!\!$} & 13 & 23 & 6 & 15\\
			 propagation of distributions via Monte Carlo & Q11 & 7 & 8 & 1 & 9\\
			 validate LPU against Monte Carlo results & Q13 & 7 & 8 & 1 & 6\\
			 LPU \& Monte Carlo for multivariate models & Q8, Q12 & 3 & 2 & 0 & 2\\
			 reporting of measurement results & Q18 & 14 & 22 & 6 & 16\\\hline
			 number of courses & & 15 & 24 & 6& 16\\
			\end{tabular}
			\end{indented}
		\end{table*}

  In addition to this range of uncertainty courses offered by and for single NMIs, EURAMET e.V.\ and the BIPM promote knowledge transfer in general. The respective websites \cite{EurametKT,BIPME-Learn} offer some training material and courses including measurement uncertainty (see section~\ref{subs:EMNs}). The dedicated activity MU Training of the European Metrology Network for Mathematics and Statistics provides support for teachers of uncertainty and training material such as videos explaining uncertainty aspects and overviews surveying uncertainty courses, software and examples (see \cite{MUTrain} and section~\ref{subs:Teach}).
	
 \subsection{Accreditation Bodies and Technical Assessors}\label{subs:Accred}

	The 109 accreditation bodies that are signatories of the ILAC MRA \cite{ILACWeb} assess and recognize the competence of conformity assessment bodies (i.e., laboratories, inspection bodies, proficiency testing providers, reference materials producers and biobanks) \cite{ILACP4}. As a consequence, worldwide confidence is established in accredited services \cite{ILACP5} and accreditation is one (core) dimension of quality infrastructure \cite{QI4SD,Keller2019}. 

  Accreditation bodies shall have the competence to assess conformity assessment bodies (c.f.\ \cite[sec.\ 1.2-3]{ILACP4}, \cite{ISO17011}), which includes assessing that these bodies competently evaluate uncertainties in calibration, when relevant in testing \cite{ISO17025}, in medical laboratories \cite[sec.\ 7.3.4]{ISO15189}, for inspections \cite[sec.\ 6.2.7-9]{ISO17020}, in proficiency testing \cite[sec.\ 4.4.5]{ISO17043}, as well as when producing reference materials \cite[sec.\ 7.13]{ISO17034} and operating biobanks \cite[sec.~6.5]{ISO20387}. 
	To do so, accreditation bodies define, evaluate and monitor competence criteria and have a process to train assessors \cite[sec.\ 6.1]{ISO17011}. In particular, they shall identify training needs and provide access to specific training.	Some larger accreditation bodies (e.g., UKAS, ACCREDIA and ANAB) offer training themselves and sometimes guidance (e.g.\ \cite{UKAS3003,COFRAC}) on evaluating uncertainty for testing and calibration laboratories.   
	
	Among the course survey \cite{CourseSurvey}, only one online-course \cite{JRCCourse} particularly includes among its audiences `auditors who need to implement and assess the estimation of measurement uncertainties [...] of reference materials'. Another course is offered by the UK accreditation body (UKAS), that also identifies `Measurement uncertainty is a key concern in all accredited calibration and testing, consequently the availability of specialist training is vitally important' [SL16]. The Italian accreditation body (ACCREDIA) provides uncertainty training during summer schools and states that its `technical officers have experience in [...] providing training courses on the standard for accreditation \cite{ISO17025} focusing on measurement traceability and the evaluation of measurement uncertainty' [PS]. In Europe, also the national accreditation bodies in Belgium (BELAC), Germany (DAkkS) and Ireland (INAB) expressed an interest in improving training on measurement uncertainty [SL17,SL27,SL29] -- mainly for laboratory staff (see section~\ref{subs:NeedCali}).
	
	In addition, a questionnaire inquiring about the interest, the (self-reported) knowledge and the preferred teaching method on 21 topics related to measurement uncertainty was prepared (see \ref{AppQuest} for details). Table~\ref{TabAccredTopics} lists the 21 topics about which technical assessors of calibration and testing laboratories among the members of the European co-operation for Accreditation (EA) and in Italy\footnote{Despite the high number of Italian participants in the questionnaire because the Italian accreditation body conducted it, the focus in this article shall be on the European (EA) participants.} were questioned. The table highlights in dark gray those topics about which the assessors of calibration laboratories reported highest knowledge, in gray medium knowledge and in white less knowledge. (On a scale from 1 to 4, average scores for EA range from 3.5 to 2.6 for high knowledge, from 2.5 to 2.0 for medium and from 1.9 to 1.5 for less, see table~\ref{TabAccredTopics}. These scores were chosen such that each column of the table contains 9-10 topics in the highest level, 6-7 in the medium level and 4 in the lowest level.)
		For basic concepts, input quantities, combined and expanded uncertainty (Q1-7), as well as reporting uncertainties (Q18) and statement of conformity (Q21) generally high knowledge is reported. For Italian assessors, high knowledge is reported also for Q16 (uncertainty from ILC/PF).	
	
	Technical assessors for testing laboratories generally report slightly lower levels of knowledge. (Average scores for EA range from 3.0 to 2.32 for high knowledge, from 2.31 to 2.0 for medium and from 1.7 to 1.5 for less knowledge). Table~\ref{TabAccredTopics} highlights these levels of knowledge. For basic concepts (Q1-3), input quantities (Q4), expanded uncertainty (Q7), reporting uncertainties (Q18), statement of conformity (Q21) and uncertainty from validation data (Q15) generally high knowledge is reported. While EA's technical assessors report high knowledge also for Q5 (combined uncertainty), Italian assessors do so for Q16 (uncertainty from ILC/PF).
		
 \subsection{Calibration Laboratories}\label{subs:Cali}

  Accredited laboratories must comply with ISO/IEC 17025 \cite{ISO17025} when performing calibration (see also \cite{ILACP14}) and testing within their accredited scope \cite{ILACR6}, and thus establish and maintain metrological traceability of their measurement results \cite{ISO17025}. The latter requires identifying the contributions to uncertainty \cite[Note 3 in 7.6]{ISO17025} and, for calibration, shall comply with the GUM \cite{ILACP14}. EA-4/02 \cite{EA-4/02} implements the GUM for calibrations and is mandatory for EA members.

 Consequently, the more than 11\,000 calibration laboratories \cite{ILACWeb} accredited by ILAC signatories, shall have the competence for evaluating measurement uncertainty in compliance with the GUM when stating it on calibration certificates \cite{ILACP14}. Calibration laboratories thus require an understanding on evaluating and reporting uncertainty. 

  Among the survey of courses on or including measurement uncertainty (see \ref{AppSurvey} and \cite{CourseSurvey, Penn2023}), there are 24 courses aimed at audiences that include employees from calibration and testing laboratories. Out of these 24 courses$^1$, 18 are dedicated to uncertainty, 5 to calibration, three to metrology and one to reference materials. Table~\ref{TabSurv} gives the numbers of these courses that teach different technical aspects of measurement uncertainty. 
	All of these courses provide some sort of certification. All but one course use at least one example and 13 use software in class. A third of the courses is given physically, a third physically and online, and a third online in pre-recorded form. 
	
 Calibration laboratories among the members of EA and in Italy were also interviewed with the questionnaire on interest, (self-reported) knowledge and preferred teaching method on 21 topics related to measurement uncertainty 
	 (see \ref{AppQuest} and table~\ref{TabAccredTopics}). 	The topics are generally grouped to the same knowledge level as for the technical assessors, but the laboratories report slightly lower knowledge scores (by about 0.2 to 0.4 score points).
		
		\begin{sidewaystable}
		\vspace{5ex}
			\caption{\label{TabAccredTopics}Questionnaire topics about which EA and Italian accredited calibration and testing laboratories and their assessors indicated their knowledge (`know') and interest (`inter'). The values give the average score and the color highlights the level (dark gray = highest level, light gray = medium level, white = lower level) of reported knowledge and interest (see \ref{AppQuest} and text for details). \emph{Emphasized} topics coincide in large parts with topics also included in the course survey~\cite{CourseSurvey}.}		
			\begin{indented}\item[]\hspace*{-8em}
				\begin{tabular}{@{}p{5em}p{0.31\textwidth}rp{1.3em}p{1.5em}p{1.6em}p{1.5em}|p{1.6em}p{1.5em}p{1.6em}p{1.5em}|p{1.6em}p{1.5em}p{1.6em}p{1.5em}|p{1.6em}p{1.5em}p{1.6em}p{1.4em}@{}}					& & & \multicolumn{4}{c|}{accredited calib.\ labs} & \multicolumn{4}{c|}{techn.\ assessors} & \multicolumn{4}{c|}{accredited test labs} & \multicolumn{4}{c}{techn.\ assessors}\\
				  & Questionnaire topic && \multicolumn{2}{c}{EA} & \multicolumn{2}{c|}{Italy} & \multicolumn{2}{c}{EA} & \multicolumn{2}{c|}{Italy}  & \multicolumn{2}{c}{EA} & \multicolumn{2}{c|}{Italy} & \multicolumn{2}{c}{EA} & \multicolumn{2}{c}{Italy}\\
				& & & \mbox{\hspace{-0.3em}know} & inter & know & inter & know & inter & know & inter& know & inter & know & inter & know & inter & know & inter\\\hline
					\multirow{3}{6em}{Basic math. \& metrolog.\ concepts}  & \emph{Mathem.\ elements for evaluating uncertainty} & Q1 &\Cellhi 2.81 & \Cellhi 3.60 & \Cellhi 2.65 & \Cellhi 3.37 & \Cellhi 2.93 & \Cellhi 3.74 & \Cellhi 3.12 & \Cellhi 3.60 & \Cellhi 2.53 & \Cellhi 3.33 & \Cellhi 2.57 & \Cellhi 3.23 & \Cellhi 2.68 & \Cellhi 3.33 & \Cellhi 2.74  & \Cellmed 3.31 \\ 
					& \emph{Probability and statistics elements} & Q2 & \Cellhi 2.53 & \Cellhi 3.50 & \Cellhi 2.46 & \Cellhi 3.40 & \Cellhi 2.93 & \Cellhi 3.67 & \Cellhi 3.00 & \Cellhi 3.48 &  \Cellhi 2.44 & \Cellhi 3.21 &\Cellhi 2.44 & \Cellhi 3.24 & \Cellhi 2.63 & \Cellhi 3.24 & \Cellhi 2.55 & \Cellmed 3.36\\
					& \emph{Fundamental concepts of metrology} & Q3 &\Cellhi 3.21 & \Cellhi 3.67 & \Cellhi 3.17 & \Cellhi 3.31 & \Cellhi 3.42 & \Cellhi 3.72 & \Cellhi 3.64 & \Cellhi 3.48 & \Cellhi 2.80 & \Cellhi 3.47 & \Cellhi 2.85 & \Cellhi 3.50 & \Cellhi 2.94 & \Cellhi 3.51 & \Cellhi 3.06 & \Cellhi 3.57  \\\hline
					\multirow{8}{6em}{Propagating uncertainties (GUM approach)} & \emph{Evaluation of type A \& B uncertainty \mbox{components\hspace{-1.2em}}} & Q4 &\Cellhi 2.79 & \Cellhi 3.66 & \Cellhi 2.54 & \Cellhi 3.47 & \Cellhi 3.00 & \Cellhi 3.79 & \Cellhi 3.20 & \Cellhi 3.64 & \Cellhi 2.22 & \Cellhi 3.13 & \Cellhi 2.28 & \Cellmed 3.20 & \Cellhi 2.41 & \Cellmed 3.14 & \Cellhi 2.70 & \Cellhi 3.44\\
					& \emph{Combined stand.\ uncertainty for uncorr.\ inputs} & Q5 & \Cellhi 2.60 & \Cellhi 3.47 & \Cellhi 2.36 & \Cellhi 3.29 & \Cellhi 2.95 & \Cellhi 3.67 & \Cellhi 3.20 & \Cellhi 3.60 & \Cellhi 2.18 & \Cellmed 3.05 & \Cellmed 2.07 & \Cellmed 3.12 & \Cellhi 2.32 & \Cellmed 3.13 & \Cellmed 2.35 & \Cellmed 3.32\\
					& Combined stand.\ uncertainty for correlated \mbox{inputs\hspace{-1.3em}} & Q6 & \Cellhi 2.37 & \Cellhi 3.48 & \Cellhi 2.24 & \Cellhi 3.42 & \Cellhi 2.63 & \Cellhi 3.72 & \Cellhi 2.76 & \Cellhi 3.52 & \Cellmed 2.16 & \Cellmed 3.06 & \Cellmed 2.07 & \Cellmed 3.11 & \Cellmed 2.29 & \Cellmed 3.01 & \Cellmed 2.39 & \Cellmed 3.35\\
					& Expanded uncertainty ($U$) \& coverage factors \mbox{($k$)\hspace{-0.8em}} & Q7 & \Cellhi 2.84 & \Cellhi 3.57 & \Cellhi 2.83 & \Cellhi 3.38 & \Cellhi 3.19 & \Cellhi 3.79 & \Cellhi 3.36 & \Cellhi 3.48 & \Cellhi 2.54 & \Cellhi 3.29 & \Cellhi 2.61 & \Cellhi 3.36 & \Cellhi 2.66 & \Cellhi 3.24 & \Cellhi 2.89 & \Cellhi 3.47\\
					 & \emph{Applying multivariate measurement models} &  Q8 & \Cellmed 1.98 & \Cellmed 3.24 & \Cellmed 1.66 & 2.90 & \Cellmed 2.12 & 3.51 & \Cellmed 2.16 &\Cellmed 3.28 & \Cellmed 1.84 & \Cellmed 2.82 & \Cellmed 1.65 & \Cellmed 2.83 & \Cellmed 2.07 & \Cellmed 2.96 & \Cellmed 1.85 & 2.99\\
					& Theoretical or empirical measurement models & Q9 & \Cellmed 2.11 & \Cellmed 3.24 & \Cellmed 1.99 & \Cellmed 3.11 & \Cellmed 2.28 & \Cellmed 3.60 & \Cellmed 2.52 & \Cellmed 3.44 & \Cellmed 2.04 & \Cellmed 2.91 & \Cellmed 1.93 & \Cellmed 3.05 & \Cellmed 2.21 & \Cellmed 3.00 & \Cellmed 2.11 & \Cellmed 3.15 \\
					& Fitness for purpose and target uncertainty & Q17 & \Cellmed 2.04 & \Cellmed 3.23 & \Cellmed 2.12 & \Cellmed 3.29 & \Cellmed 2.47 &\Cellmed 3.56 & \Cellmed 2.72 & \Cellhi 3.48 & \Cellmed 2.05 & \Cellmed 3.08 & \Cellmed 2.23 & \Cellhi 3.43 & \Cellmed 2.31 &\Cellhi 3.19 & \Cellmed 2.52 & \Cellhi 3.57\\
					& \emph{Reporting measurement results} & Q18 & \Cellhi 2.86 & \Cellhi 3.53 & \Cellhi 2.90 & \Cellmed 3.28 & \Cellhi 3.19 & \Cellhi 3.72 & \Cellhi 3.32 & 3.25 & \Cellhi 2.66 & \Cellhi 3.42 & \Cellhi 2.84 & \Cellhi 3.47 & \Cellhi 2.83 & \Cellhi 3.48 & \Cellhi 3.07 & \Cellhi 3.54 \\
					Conformity & Statements of conformity to specifications	& Q21 & \Cellhi 2.27 & \Cellhi 3.34 & \Cellhi 2.45 & \Cellhi 3.49 & \Cellhi 2.65 & \Cellhi 3.74 & \Cellhi 2.84 & \Cellmed 3.44& \Cellhi 2.26 & \Cellhi 3.20 & \Cellhi 2.39 & \Cellhi 3.43 & \Cellhi 2.45 & \Cellhi 3.29 & \Cellhi 2.80 & \Cellhi 3.65  \\\hline				
					\multirow{4}{6em}{Uncertainty evaluation for specific data} & Least squares method applied to metrology & Q10 & \Cellmed 2.18 & \Cellmed 3.29 & \Cellmed 2.07 & \Cellmed 3.24 & \Cellmed 2.49 & \Cellmed 3.65 & \Cellmed 2.68 & \Cellhi 3.52 & \Cellmed 1.91 & \Cellmed 2.76 & \Cellmed 2.07 & \Cellmed 2.98 & \Cellmed 2.19 & \Cellmed 2.92 & \Cellmed 2.24 & \Cellmed 3.15\\
					& Uncertainty based on methods validation data & Q15 & \Cellmed 1.96 & \Cellmed 3.27 & \Cellmed 1.83 & \Cellhi 3.30 & \Cellmed 2.00 & \Cellmed 3.53 & \Cellmed 2.25 & \Cellmed 3.28& \Cellhi 2.34 & \Cellhi 3.36 & \Cellhi 2.45 & \Cellhi 3.49 & \Cellhi 2.51 & \Cellhi 3.33 & \Cellhi 2.72 & \Cellhi 3.56\\ 
					& Uncertainty based on ILC/PT data \& \mbox{experience\hspace{-0.8em}} & Q16 &\Cellmed 2.04 & \Cellmed 3.30 & \Cellhi 2.28 & \Cellhi 3.61 & \Cellmed 2.16 & \Cellmed 3.58 & \Cellhi 3.00 & \Cellhi 3.52 & \Cellmed 2.15 & \Cellhi 3.21 & \Cellhi 2.33 & \Cellhi 3.47 & \Cellmed 2.30 & \Cellmed 3.18 & \Cellhi 2.61 & \Cellhi 3.55 \\
					& Uncertainty for sampling & Q20 &\Cellmed 1.82 & 2.89 & \Cellmed 1.65 & 2.84 & \Cellmed 2.16 & 3.47 & 2.00 & 3.04 & \Cellmed 1.99 & \Cellmed 3.12 & \Cellmed 1.80 & \Cellmed 3.10 & \Cellmed 2.26 & \Cellhi 3.26 & \Cellmed 2.11 & \Cellhi 3.48 \\\hline	
					\multirow{3}{6em}{Propagating distributions (Monte \mbox{\hspace{-0.25em}Carlo)\hspace{-1em}}}& \emph{Single measurand (Univariate model)} & Q11 & 1.42 & \Cellmed 3.00 & 1.60 & \Cellmed 3.09 & 1.93 & \Cellmed 3.63 & \Cellmed 2.08 & \Cellmed 3.28 & 1.45 & 2.38 & 1.44 & 2.63 & 1.62 & 2.63 & 1.54 & 2.89\\
					& \emph{More measurands (Multivariate model)} & Q12 & 1.37 &	2.95 & 1.49 & 2.94 & 1.72 & 3.51 & 1.84 & 3.24 & 1.37 & 2.31 & 1.40 & 2.57 & 1.55 & 2.55 & 1.42 & 2.85 \\
					&  \emph{vs.\ GUM approach for uncertainty evaluation}& Q13 & 1.44 & 2.93 & 1.50 & \Cellmed 3.11 & 1.84 & \Cellmed 3.60 & 1.92 & \Cellmed 3.36 & 1.42 & 2.50 & 1.36 & 2.70 & 1.67 & 2.72 & 1.43 & 3.06\\[0.25em] 
					Bayes \mbox{appr.\hspace{-0.25em}} & Alternative methods for uncertainty & Q14	& 1.44 & 2.86 & 1.30 & 2.79 & 1.49 & 3.26 & 1.64 & 3.20 & 1.34 & 2.46 & 1.35 & 2.62 & 1.63 & 2.66 & 1.58 & \Cellmed 3.06\\[-70ex]
				\end{tabular}
				\end{indented}
		\end{sidewaystable}

		Tables~\ref{TabSurv} and~\ref{TabAccredTopics} coherently show that LPU as well as its input and reporting (Q4,5,18) are frequently taught (i.e.\ in more than 90 \% of the courses) and there is high knowledge about it for European and Italian calibration laboratories.

 \subsection{Testing Laboratories}\label{subs:Test}

  Accredited laboratories must comply with ISO/IEC 17025 \cite{ISO17025} when performing testing within their accredited scope \cite{ILACR6}, and thus establish and maintain metrological traceability of their measurement results \cite{ISO17025}. The latter requires identifying the contributions to uncertainty \cite[Note 3 in 7.6]{ISO17025}.

  Analogous to calibration laboratories, also the 65\,000 testing laboratories \cite{ILACWeb} accredited by ILAC signatories shall evaluate, or at least make an estimation of, the uncertainty for their measurement results \cite[sec.\ 7.6.3]{ISO17025}. However, ILAC G17 \cite[clause 3]{ILACG17} also recognizes documents alternative to the GUM. Testing laboratories thus require an understanding on evaluating and reporting uncertainty, although at different levels than calibration laboratories. 

 Among the survey of courses on or including measurement uncertainty, there are 24 courses aimed at audiences that include employees from calibration and testing laboratories (see section~\ref{subs:Cali}). That is, the survey gives some information on uncertainty training for testing laboratories, but it does not separate this audience from calibration laboratories.

 Testing laboratories among the members of EA and in Italy were interviewed with the questionnaire on interest, (self-reported) knowledge and preferred teaching method on 21 topics related to measurement uncertainty 
	 (see \ref{AppQuest} and table~\ref{TabAccredTopics}).  
 Testing laboratories generally report slightly lower levels of knowledge than their assessors (by about 0.1 to 0.3 score points) and than calibration laboratories (by up to 0.4 points). The topics are generally grouped to the same knowledge level as for their technical assessors (see section~\ref{subs:Accred}).

	 In addition, tables~\ref{TabSurv} and~\ref{TabAccredTopics} coherently show that input quantities and reporting uncertainties (Q4,18) are frequently taught ($>$90 \%) and there is high knowledge about it for European and Italian testing laboratories. 
		
 \subsection{Legal Metrology Authorities and their Organisations}\label{subs:Legal}	
  Legal metrology aims to ensure trust and fairness in measurements covering trade, as well as to protect health, safety and the environment \cite{OIMLD1,Keller2019}. 
Within this scope, national regulations lay down measurement-based requirements and requirements for measuring instruments or systems \cite{OIMLBroch}.  
The International Organization of Legal Metrology (OIML) with its 64 member states and 63 corresponding states promotes the global harmonization of legal metrology laws, and provides its members with guidance on their national legislation \cite{OIMLD1}.  In particular, OIML D 1 \cite[sec.\ 6.5, element no.\ 10]{OIMLD1} recommends that measurement results should be traceable when covered by regulations, performed to control regulated prepackages and provided by regulated instruments, and non-fulfilment should be an offence \cite[element no.\ 29]{OIMLD1}. Requirements on regulated measurements themselves ordinarily include required measurement uncertainties \cite[sec.\ 6.5.1, element no.\ 17]{OIMLD1}. In all these cases, uncertainties should be established following the GUM \cite{JointDeclar}, \cite[art.\ 17]{OIMLD1}. In addition, conformity assessment procedures to enforce regulations are recommended to follow OIML guidance \cite[sec.\ 6.6]{OIMLD1}, which includes the recent document \cite{OIMLG19} on how to take uncertainty into account in conformity decisions in legal metrology. 

Consequently, local legal metrology authorities, which implement legal controls, conduct surveillance inspections and verifications of instruments and prepackages, and accept or reject them \cite[sec.\ 3.2.4]{OIMLD1}, require an understanding of the importance and evaluation of uncertainty for regulated measurements. See \cite{OIMLD14} for training content for staff in verification offices. In addition, national legal metrology authorities which may or may not be part of a country's NMI and develop metrological controls, study requirements, calibration and test equipment, carry out or supervise type evaluation and provide training in legal metrology \cite[sec.\ 3.2.3]{OIMLD1}, require an understanding of the importance and evaluation of uncertainty in legal metrology. Furthermore, the central government authority that is in charge of the national metrology policy and coordinates metrology related government actions requires an understanding of the importance of uncertainty \cite[sec.\ 3.2.1]{OIMLD1}. An understanding of the concept of uncertainty is required also by secretariats, conveners and members of technical committees, subcommittees or project groups at OIML \cite{OIMLG19} and regional legal metrology organisations such as WELMEC.

	The European Cooperation in Legal Metrology, WELMEC, had 9 active working groups in 2022, whereof 8 were interviewed on their understanding and their need for training on measurement uncertainty (see \ref{AppWelm}). Several working groups promote aspects of uncertainty in guidelines, such as Guide 4.2 \cite{WELMEC4.2}, 6.9 \cite{WELMEC6.9}, and to some extent also 13.1 \cite{WELMEC13.1}. Guide 8.10 \cite{WELMEC8.10} recommends sampling plans. In general, WELMEC guides are aimed at various conformity assessment activities for the legal control of regulated measuring instruments [SL15]. The WG convenors are usually NMI employees and stated in the interview to have themselves appropriate knowledge on measurement uncertainty.

	Among the survey of courses on or including measurement uncertainty (see \ref{AppSurvey} and~\cite{CourseSurvey, Penn2023}), there are 6 courses aimed at audiences from verification authorities (cf.\ table~\ref{TabSurv}). These 6 courses are offered either by one of three NMIs or by a federated academy for legal metrology. They include$^1$ 4 courses dedicated to uncertainty and one each to metrology experiments and to gas pump verification. Table~\ref{TabSurv} gives the number of these courses that teach different technical aspects of measurement uncertainty. All but one course use software in class. The same courses provide at least one example each. Two courses are available also online and one online only (in prerecorded form). All courses provide certification, half of them after an exam.

 \subsection{Students, Lecturers and Researchers at Universities}\label{subs:Uni}

 The evaluation of uncertainty is an essential part of courses on measurement data in University degree programs in metrology and physics \cite{AAPT2014,Wan22,BorWa20}. Aspects of uncertainty are taught also in degree programs in engineering \cite{Chimeno2005}, chemistry \cite{Kusel2008,ChemTart}, biology \cite{Schanning2019,Whit08}, and medicine \cite{MedizLeipz}, especially laboratory medicine, as well as in related fields. 
Literature shows that state-of-the-art teaching of measurement uncertainty follows the concepts of the GUM. At the same time it shows that  
not all courses do so yet \cite{Schanning2019}. In addition to students and lecturers, also researchers whose work is based on measurement results are in need of using, reporting or evaluating uncertainties, and it is their responsibility to use pertinent methods to do so.

  Among the survey of courses (see \ref{AppSurvey} and \cite{CourseSurvey, Penn2023}), there are 16 courses aimed at audiences that include students, lecturers and/or researchers at universities. Out of these 16 courses$^1$, 9 are dedicated to uncertainty, 6 to measurement or metrology, one to testing and certification and one to reference materials. Table~\ref{TabSurv} gives the number of these courses that teach different technical aspects of measurement uncertainty. All of these courses provide some sort of certification and use at least one example in class. Ten courses use software in class for propagating uncertainty or calculating uncertainty budgets. Of the courses 5 are given physically, 7 physically and online and 4 online only (most in pre-recorded form). 
		
	Considering the number of universities in Europe and the number of degree programs they offer, the survey of uncertainty courses may not give a representative view. Particularly those universities or departments without NMI contacts or not teaching GUM concepts may be underrepresented in the survey, considering that 9 of the 16 courses at universities included in the survey are taught by NMI staff.
	
In addition to the survey, research gives insights into the content of uncertainty courses at universities and schools. For example, \cite[p.\ 34]{Kok22} summarizes that `measurement uncertainties is traditionally introduced at universities during laboratory courses \cite{AllBu03,HuZwick21,MoeRu20,Pollard2021}). [...] Often, these laboratory courses are accompanied by theoretical (statistics) courses that introduce the topic of measurement uncertainties \cite{HuZwick21,Pollard2021,SereJour93,VolAll08}.' In secondary education, however, `measurement uncertainty is a topic that is often neglected' \cite[p.\ iii]{Kok22}.

 \subsection{Different Metrology Fields}\label{subs:EMNs}

  There are 12 European Metrology Networks (EMNs) that represent the measurement science community in different fields. 
	These EMNs analyse and address the European and global metrology challenges, and uncertainty has been explicitly identified as a challenge by 5 of these EMNs already (see \ref{AppEMN} for details).
	In particular, the EMN
	\begin{itemize}
		\item Mathmet identified the foundational topic `Data Analysis and Uncertainty Evaluation', as well as the demand for research on uncertainty to support the strategic topics of `Artificial Intelligence and Machine Learning' and `Computational Modelling and Virtual Metrology' (see \cite{MathmetSRA} and section~\ref{subs:NeedEmerg} for details).
		\item Climate and Ocean Observation identified uncertainty as a general metrology challenge, particularly for in situ observations of essential climate variables in the land domain, for observations of the ocean and for remote sensing \cite{COOExecSummary}. 
		\item Advanced Manufacturing identified requirements for uncertainty particularly in the cross-cutting topics of intelligent product design, advanced materials and smart manufacture and assembly. The EMN envisages instruments or their digital twins to provide measurement results including uncertainties, and AI algorithms to be validated and `calibrated' for uncertainty determination \cite{AdvManuSRA}. 
		\item Radiation Protection's vision is that `quality assurance including measurement traceability to the SI system is available for all measurements in the respective exposure situation addressed under the European legislation.' For this, they aim to support with reliable data including uncertainties \cite{RadProtVision}.
		\item Smart Electricity Grids identified `grid monitoring and data analytics' as one of 8 themes with particular metrological relevance to smart grids, and for this theme identified the requirement to develop `big data analytics and visualisation platforms with adequate evaluation of measurement uncertainty' and `machine learning algorithms for short-term load forecasting' \cite{SmartEGSRA}. 
		\item	Laboratory Medicine's mission is to provide metrological traceability of in vitro diagnostics, which is important in health care, and for devices regulated in EU 2017/746 \cite{TraceLabMed}. Uncertainty is not mentioned explicitly, but is a prerequisite for traceability.\end{itemize} 

EMNs were founded only in 2019 or later, such that at the beginning of 2024,
	the strategic research agenda of 6 EMNs were not yet available or did not mention the keyword `uncertainty' (c.f.\ \ref{AppEMN}). None of the EMNs has published information on the state of the art of measurement uncertainty training in their field (yet). Nevertheless, some EMNs identified uncertainty training among their priorities, see section~\ref{subs:NeedEMNs}.
	
 EURAMET advertises and archives events \cite{EurametEvents}  
 like trainings, workshops and courses that have been organized in different metrology fields. 12 events containing the keyword `uncertainty' were organized from 2017 to 2024, either in relation to projects within EURAMET's research programmes such as EMPIR, as EURAMET capacity building activity,  by bodies in close relation to EURAMET or by the EMN Mathmet. These events comprise 5 workshops, 6 training courses, and one tutorial, all explicitly dedicated to the teaching of measurement uncertainty topics, either at a general level or for a specific application (i.e.\ spectral data, volatile organic compounds measurements, chemical analysis and sampling, flow measurements, electrical power and energy, mass calibration and volume measurements). Among the 48 events containing the more generic keyword `training', 10 further events included the evaluation of measurement uncertainty as a topic in their agenda. These trainings were all provided in the last two years and by similar types of organizers as the dedicated uncertainty events above. All these events confirm the cross-disciplinary nature of measurement uncertainty and the transversality of the need for its comprehension, modelling and evaluation across many different metrology fields.

 \subsection{Teachers of Measurement Uncertainty}\label{subs:Teach}

In general, following Shulman’s model of professional knowledge of teachers \cite[p.\ 8]{Shul87}, the three dimensions content knowledge, general pedagogical knowledge and content specific pedagogical knowledge (i.e.\ knowledge on how to best teach a certain topic) are needed to successfully teach a topic. When teaching uses digital tools, technological knowledge is needed in addition and overlaps with all three dimensions mentioned above \cite{KoeMi05}. Furthermore, teaching is facilitated by a sufficient number of teachers as well as suitable material that addresses concepts and supports learning. This section focuses on the state of the art related to providing measurement uncertainty training which is common to the community of teachers, and points to available resources for teaching measurement uncertainty. Section~\ref{subs:NeedTeach} will point to further needs.

In the case of measurement uncertainty, content knowledge implies a thorough understanding of the concepts and limitations of measurement uncertainty as well as practical competence in evaluating, using and reporting measurement uncertainty as described in the GUM suite of documents \cite{JCGM100,JCGM101,JCGM102,JCGM106,GUM1,GUM6} and in research. In order to design training tailored to a specific audience, awareness on audience needs, expectations and prior training (with respect to metrology, mathematics, etc.) is also required to select relevant content, suitable examples, software etc. 
 In the subsections of section~\ref{sec:stateofart} and~\ref{sec:needs}, this specific background was described for each community and will not be repeated here.

The second and third of the three dimensions needed for successful teaching are general pedagogical skills, and pedagogical content knowledge specific to measurement uncertainty. The little research available on the latter, gives insights, for example, into learning problems \cite{SereJour93},  pre- or misconceptions \cite{LubCa01} and validated assessment tools (e.g.\ \cite{Schul22, VigGePoll23}) to monitor the progress of trainees from purely knowing about the existence of measurement uncertainty towards its handling, assessment, and finally to its conclusiveness \cite{Priem18}. In addition, \cite[p.\ 34-5]{Kok22} and \cite{Hein12} conclude that the most important aspect to successfully teach uncertainties is a concept-based approach and an emphasis on the underlying principles rather than statistical and calculational procedures alone. Training on general pedagogical skills as well as dedicated training centers may support teachers to develop courses (e.g.\ \cite{MUTPresNPL}) or skills that often were not part of their education.

Practical experience was exchanged between teachers of measurement uncertainty at a recent workshop \cite{SkillsWork22}, and emphasized the importance of examples (e.g.\ an annotated template of a simple uncertainty budget) to illustrate the content of training and to learn by example, as well as the importance of (interactive) exercises, quizzes and other practical work to transfer higher levels of cognition (c.f. \cite{Krath02}). Knowledge transfer may also be improved by interactions between trainees as well as a follow-up contact and consultations between expert and trainee after the course content was implemented. On the other hand, trainee feedback may support the improvement of future courses. In addition, general pedagogical knowledge on methods like blended learning and flipped classroom \cite{FlipClass} allow to implement more individual learning and more time for discussions in class. Exchange of good practices and experiences between teachers of uncertainty increases their capabilities; for example, the MU Training activity may serve as a reference point for uncertainty training and improve pedagogical content knowledge as well as pedagogical knowledge. Among others, the activity organized a workshop between non-professional teachers (see \cite{SkillsWork22} and \cite{MUTTeachers}), and it set up a framework to mutually attend courses within its consortium.

	The context in which uncertainty training is given differs substantially. The course survey \cite{CourseSurvey} provides insights into organizing bodies, into training frameworks such as metrology, calibration, different Master, Bachelor degrees or PhD programmes. Furthermore, the courses cover different metrology areas and have different frequencies, durations and languages.

In addition to the professional knowledge of teachers, Shulman \cite[p.\ 9]{Shul87} pointed to didactically sound material and tools to support teachers. These include tailored curricula (c.f.\ sections~\ref{sec:stateofart} and~\ref{sec:needs}), but also examples and software enhance courses on uncertainty that may bridge the gap between theory and application. Surveys on both \cite{ExSurvey,SurveySoftware} provide guidance for choices tailored to the audience and context. Introductory material and e-learning may help to align background knowledge of heterogeneous audiences. The MU Training activity offers the exchange of course material within its consortium. The GUM documents themselves \cite{GUM1,JCGM100,JCGM101,JCGM102,JCGM106,GUM6}, textbooks (e.g.\ \cite{Lira02,GerSo24}) and guidelines in different metrology fields support teaching uncertainty. Video material may also support teachers, and is available on certain aspects of uncertainty \cite{Videos}. A Digital Learning Environment on Measurement Uncertainties including videos and practice problems for secondary school students was developed \cite{KokDLE} and assessed in \cite{Kok22}. Dedicated e-learning courses are also offered by CEM in Spanish, by LNE in French, and by JRC and NPL in English (see \cite{CourseSurvey}). More generally, educational technology has been shown to lead to positive effects on learning outcomes if used intentionally \cite{Seuf21}. Technological tools are abundant and some multi-purpose ones are listed \cite{FlipClass}.

\section{The training needs of each community}\label{sec:needs} 
For each of the communities of NMIs, accreditation bodies, calibration and testing laboratories, legal metrology, universities and different metrology fields, this section will identify separate needs for their training on uncertainty. While communities may also have needs for tailored guidance, examples or software on uncertainty, those needs will be touched upon only when related to training.  

Training providers, such as the European NMIs, are usually interested in how to better deliver their training. If particular training needs are not specific to any of the audiences, they are deferred to section~\ref{subs:NeedTeach} which summarizes the teachers' needs. In addition, section~\ref{subs:NeedEmerg} will identify needs that may emerge due to new developments in metrology. 

Section~\ref{sec:Concl} will then summarize the needs identified in this section and will give recommendations as well as an outlook.

 \subsection{NMIs, DIs and other Signatories of the CIPM MRA}\label{subs:NeedNMI}
	
	The survey on uncertainty courses in Europe \cite{CourseSurvey} highlighted a lack of training on multivariate models \cite{JCGM102} and thus on the calculation of correlation between multiple measurands -- even though these correlations should always be evaluated when measurands depend on common input quantities. Also the rather small number of courses dedicated to propagating distributions may imply a need (c.f. table~\ref{TabSurv}).
	
	Furthermore, some of the smaller NMIs aim to extend their offer on uncertainty training because they currently do not teach courses or would like to serve new audiences [PS]. 
		Emerging NMIs stated a need for training on uncertainty methods for themselves [SL20, on MC] and a need for uncertainty trainers to teach their stakeholders. A DI would like to strengthen its teaching on uncertainty due to retiring teachers and with regard to aspects such as sampling, calibration, testing, binary or ranked tests and conformity assessment [SL22].	

 \subsection{Accreditation Bodies and Technical Assessors}\label{subs:NeedAccred}
  The current sources of information provide little insight into training of accreditation body personnel. Neither the course survey \cite{CourseSurvey} nor the 5 accreditation bodies that are partners or stakeholders of the MU Training activity include training explicitly tailored to this audience. The authors suspect that this community 
	attends training that is mainly directed at other audiences, such as NMI or laboratory staff.
	
  For technical assessors, the questionnaire described in section~\ref{subs:Accred} and \ref{AppQuest}, also inquired about their level of interest in 21 uncertainty topics (see table~\ref{TabAccredTopics}). In general, the level of interest is similar to the level of reported knowledge
for the topics, but the average interest score is higher and less spread than the average knowledge score. For example, for assessors of EA calibration laboratories average scores range from 3.8 to 3.65 for high interest, from 3.65 to 3.53 for medium interest and from 3.51 to 3.26 for less interest. This outcome may be interpreted as a general lack of structured training on uncertainty topics. 

 Let us consider topics which raise even more interest than knowledge relative to other topics and which could thus be taught more for technical assessors. These topics are Q11 (univariate Monte Carlo) and Q13 (GUM vs.\ Monte Carlo) for assessors of EA calibration laboratories. For assessors of testing laboratories, topics Q17  (fitness for purpose of uncertainty) and Q20 (uncertainty for sampling) raise more interest than knowledge. Among Italian assessors, reported knowledge and interest seem to diverge on more topics.
 For assessors of EA testing laboratories, 
 the prerequisites for topics of high knowledge and interest (Q7,18,21), surprisingly, do not stimulate equally high interest. That is, interest and partially also knowledge is lacking for input and combined uncertainties (Q4-6) and for modelling (Q8-9). This outcome may be explained by the requirement in \cite[sec.\ 7.8.3.1]{ISO17025} to state uncertainties in test reports only when relevant to the test result, the customer or the conformity statement. (For example, stating conformance based on simple acceptance with an uncertainty or uncertainty sources limited by a documented test method \cite[sec.\ 8.2.4, 8.2.5]{JCGM106} may avoid explicitly evaluating the uncertainty.) Assessors generally reported less interest for Q12 (multivariate Monte Carlo) and Q14 (Bayesian methods). While for assessors of EA calibration laboratories also interest in Q8 (multivariate LPU) and Q20 (uncertainty for sampling) is low, for testing laboratories it is also low for Q11 (univariate Monte Carlo) and Q13 (GUM vs.\ Monte Carlo).

 For each of the topics in the questionnaire, the respondents were also asked about their preferred teaching method among the options `Theoretical', `Exercises illustrated by the teacher', `Exercises carried out by the attendee' and `Use of dedicated software'. For all topics, `Exercises illustrated by the teacher', is the (or among the) most preferred teaching method for technical assessors. One could interpret this as the general demand for more exercises in uncertainty courses (c.f.\ sections~\ref{subs:MU} and~\ref{subs:Teach} for the importance and a caveat of exercises and examples).
	
 \subsection{Calibration Laboratories}\label{subs:NeedCali}
	Compared to NMIs, it is expected that calibration and testing laboratories generally have less statistical and numerical competence when evaluating uncertainties ([SL6] and \cite{MUTPresTesto}). 
		
\noindent Calibration laboratories offering metrology training expressed the need
	\begin{itemize}
		\item for a common comparable approach and clear structured educational systematic for teaching measurement uncertainty [SL8]
		\item for comparable uncertainty training and expertise [SL11], and
		\item to better transfer the mathematical basics into practice in calibration and testing [SL4].
	\end{itemize}
	A standardization body expressed the need for uncertainty training that is particularly directed at European calibration and testing laboratories and assumes that the recent edition of ISO 17025 poses challenges to this audience that require training [SL6], e.g.\ conformity statements that account for uncertainty. 
In addition, national accreditation bodies 
\begin{itemize}
		\item are `interested in teaching that is targeted at the laboratory practitioner level including basic measurement equations, methods to recognise and address correlation, and calculation, reporting and use (importing) of measurement uncertainty within GUM-LPU in the presence of dominant type B uncertainties' [SL16]
		\item identified the need to `improve the trainings in this technical field [authors' note: of determining uncertainties in testing and calibration performed by accredited laboratories] and with that the competence of all participating laboratory staff' [SL17]
		\item stated the importance of `setting up training to raise awareness about measurement uncertainties and to ensure the deep understanding of their concepts' for all laboratories that apply for accreditation [SL28], and 
		\item identified the need for `common, best quality material to fill the gap by appropriate trainings targeting very different audiences' because `founding concepts are often not really understood, because drown in extensive ex-cathedra calculus' [SL28]. 
	\end{itemize}
		
	The questionnaire summarized in table~\ref{TabAccredTopics}
	asked laboratories also about their level of interest in 21 uncertainty topics. High levels of interest there largely coincide with the expectations on uncertainty training, that a training provider \cite{MUTPresTesto} observed to be selecting, modelling and describing uncertainty sources (Q4-5), establishing and extending the basic measurement model (Q9), covariance (Q6,8), Monte Carlo (Q11) and decision rules (Q7,21). In general, the level of interest is similar to the level of reported knowledge for the topics, but the average interest score is higher and less spread than the average knowledge score (see table~\ref{TabAccredTopics}). 
	Again, this outcome may be interpreted as a general lack of training on uncertainty topics and confirms the need expressed above by standardization and accreditation bodies. In addition, topics exist that raise even more interest than knowledge relative to other topics. This is generally the case for Q11 (univariate Monte Carlo) and Q13 (GUM vs.\ Monte Carlo). Among Italian laboratories also other topics raise relatively more interest than knowledge. On the other side, uncertainty for sampling (Q20) 
	raises less interest compared to other topics.
		
	Tables~\ref{TabSurv} and~\ref{TabAccredTopics} coherently show that the propagation of distributions via Monte Carlo and the validation of LPU against Monte Carlo results are taught little (33 \%) and there is little knowledge about it for European and Italian calibration laboratories (generally in the lower third). However, the interest in both topics exists according to table~\ref{TabAccredTopics}. The multivariate extension of LPU and Monte Carlo is taught even less ($<$10~\%), and the interest is medium and low, respectively -- despite the need to account for correlations when measurands depend on common input quantities. Because calibrations often imply correlation, there may be a need to raise awareness about the importance of multivariate (joint) modelling of measurands.
				
	The preferred teaching method is, again, `Exercises illustrated by the teacher' for all topics.
	One could interpret this preference as the general demand for more exercises in uncertainty courses, which reiterates the need to better transfer knowledge into practice and for tailored courses as was expressed above by a training provider and by standardization and accreditation bodies.
		
 \subsection{Testing Laboratories}\label{subs:NeedTest}
 A manufacturer to whom product and component testing is essential, stated the need for training on how to identify sources of uncertainty, or more generally how to apply the `hard math' in practice, and the need for different levels of training for development engineers, managers and suppliers [SL23]. A testing laboratory aims to keep its knowledge up-to-date and to improve uncertainties, amongst others by being informed about uncertainty software and courses [SL32].

 More generally, table~\ref{TabAccredTopics} ranks the interest in uncertainty topics among accredited testing laboratories. Testing laboratories generally report average interest scores that are higher and less spread than the average knowledge score. 
Again, this may be interpreted as a general lack of training on uncertainty topics. 
 For testing laboratories, the prerequisites for topics of high knowledge and interest (Q7,18,21), surprisingly, do not stimulate equally high interest. That is, interest and partially also knowledge is lacking for correlated input and combined uncertainties (Q5,6) and for modelling (Q8,9).
(C.f.\ assessors of testing laboratories in section~\ref{subs:NeedAccred} for an explanation.) 
	
  Tables~\ref{TabSurv} and~\ref{TabAccredTopics} coherently show that the propagation of distributions via Monte Carlo (Q11) and the validation of LPU against Monte Carlo results (Q13) are taught little (33 \%) and there is little knowledge and little interest about it for European and Italian testing laboratories (in the lower third). The multivariate extension of Monte Carlo (Q12) is taught even less ($<$10~\%), and the interest is low.
	
	Again, the preferred teaching method is `Exercises illustrated by the teacher' for all topics,
	which could be interpreted as the general demand for more exercises in uncertainty courses.
 
 \subsection{Legal Metrology Authorities and their Organisations}\label{subs:NeedLegal}

	WELMEC stated a general shortage of teaching material and courses that are both accessible and understandable to different specialists in legal metrology [SL15], such as notified bodies or field inspectors. The MU Training activity intends to address this need by proposing a curriculum that can serve as a basis and may be adapted to particular audiences. During the interviews summarized in \ref{AppWelm}, WELMEC WG 6 stated the additional need for support to improve the Guide 6.9 \cite{WELMEC6.9}, especially on model building, and to develop software including uncertainty for the packer procedure. WG 8 stated the need  
	for conformity assessment according to JCGM 106 \cite{JCGM106}. In addition, one federated academy for legal metrology stated the need to offer more blended learning with a larger online portfolio for teaching [PS]. 

 The course survey \cite{CourseSurvey} shows a particular lack of courses on propagating distributions via Monte Carlo and on multivariate models that are accessible for legal audiences. In addition, the few courses covered by the survey highlight the need either for a more complete overview or for more courses on measurement uncertainty for the legal community. WELMEC's statement above indicates the latter need. Furthermore, when new national or European legislation or guidance is adopted, existing courses may need to be adapted.

 \subsection{Students, Lecturers and Researchers at Universities}\label{subs:NeedUni}
	The relatively low number of courses on uncertainty that are taught at universities and are included in the survey \cite{CourseSurvey} highlights the need for a more complete overview on such courses, especially among those university departments with weaker links to metrology.	
	Among the courses that are covered by the survey, a lack of training on multivariate models at universities is indicated.

 Three universities expressed interest in improving their teaching of uncertainty, e.g.\ with new training material, and in establishing new courses [SL21,33,34]. 
 In addition, they
\begin{itemize}
	\item would like to focus more on the concepts and importance of uncertainty, and would like to increase online delivery and other new digital teaching tools in curricula [SL33]
	\item aim to extend their offer to a course including uncertainties in regression and including the relevance of uncertainties for machine learning models and big data analytics [SL33]
	\item aim particularly at the calculation of sensitivity coefficients and of expanded uncertainties via LPU still accounting for distributions, correlations and a minimum number of measurements, as well as at evaluating uncertainty in calibrations (including needs for methods, models, budgets, validation), when sampling and via Monte Carlo (needing a review on software and uncertainty budget calculations) [SL34].
\end{itemize}	
A fourth university's physics department expressed interest in new teaching approaches of the GUM, in applications to basic research and wants to stay up-to-date [SL18]. A fifth university is interested in making available reference courses on the fundamentals of measurement uncertainty training to students in the fields of basic science and engineering [SL26].
 In addition, a cooperation expressed that the understanding of uncertainty is not yet fully established in university education in the area of traceability in analytical chemistry [SL1].
		
 \subsection{Different Metrology Fields }\label{subs:NeedEMNs}
 For the metrology fields of Advanced Manufacturing, Climate and Ocean Observation, Laboratory Medicine, Mathematics and Statistics, Radiation Protection and Smart Electricity Grids, uncertainty was identified as a challenge (see sections~\ref{subs:EMNs} and \ref{AppEMN}). For these fields, the EMN
\begin{itemize}
	\item Climate and Ocean Observation also states that `One of the most common requests [...] is for training in both instrumental techniques and in uncertainty analysis' \cite[p.\ 36f]{COOSRA}. Their roadmap until 2025 and 2032 aims to tailor training in uncertainty analysis for communities and to connect these communities to generic training developed for example through EMN Mathmet. In particular, uncertainty training courses or training support for observations of the ocean and remote sensing are needed. 
	\item Radiation Protection aims to provide expertise and to support their vision \cite{RadProtVision} with reliable data including uncertainties. There is also interest to interact with the EMN Mathmet [SL30]. 
	\item Advanced Manufacturing aims to leverage metrology advancements through knowledge transfer and training. They want to harmonise fundamental metrology courses, to connect national metrology training hubs, as well as to coordinate training material, course development and transfer \cite{AdvManuSRA}. Training on uncertainty is particularly important because manufactured components are accepted based on specified manufacturing tolerances, the measurement values for measurands on the components and their associated measurement uncertainties [SL13]. 
	\item Smart Electricity Grids stated, that its training focuses on metrological educational resources specific to scientific and engineering know-how related to electrical energy and power, and it is interested in extending their training offerings to courses on the fundamentals of measurement uncertainty [SL25]. 
	\item TraceLabMed and its members offer uncertainty training, e.g.\ for calibration laboratories, and stated their interest to receive feedback and to establish a core curriculum for their audiences [SL31]. 
	\item Mathmet identified \cite{MathmetSRA} the need for research on uncertainty (see section~\ref{subs:NeedEmerg} for details). During interviews, 7 members of Mathmet's Stakeholder Advisory Committee expressed interest in training activities on measurement uncertainty (mostly medium or high interest). One of these stakeholders [WELMEC e.V.] also suggested that training activities on applications to conformity decisions would be useful. Currently, the EMN's activity MU Training is improving the quality, efficiency and dissemination of measurement uncertainty training.
\end{itemize}

  Within EURAMET e.V.\ scientific and technical cooperation in different fields is organized in Technical Committees (TCs, see \cite{EurametTC} and its subpages). TC-Length members [SL7], the TC for Metrology in Chemistry and several subcommittees of the TC for Mass and Related Quantities are involved in knowledge transfer. The TC for Interdisciplinary Metrology moreover monitors training needs for early career researchers in metrology [TC-IM Early Career project group] and the TC-Thermometry does so in its field [SL5, TC-T Best practice WG objectives]. TC-Thermometry also initiates courses among NMIs and accredited laboratories, and pertaining to measurement uncertainty training is interested in exchange, to develop new and improve existing courses [SL5]. TC-Length considers enhancing the understanding of uncertainty concepts to be crucial for decisions in science, industry and legal metrology, and is interested in new training material for uncertainty courses. They particularly observe the need to improve uncertainty training in the practical realization and improvement of length and angle units, and also in the application of measurement techniques in fields ranging from nanotechnology via advanced manufacturing to long range measurements. They are highly interested in common training material and in exchange with the MU Training activity [SL7].
	
 \subsection{Teachers of Measurement Uncertainty}\label{subs:NeedTeach}
  Teachers of measurement uncertainty need knowledge of the subject, general and uncertainty-specific pedagogical knowledge, technology-related knowledge, knowledge of their audience and context of teaching, as well as appropriate teaching material and tools (c.f.\ section~\ref{subs:Teach}). Also the number of qualified teachers needs to be in line with the demand. Sections~\ref{subs:NeedNMI} to~\ref{subs:NeedEMNs} collected evidence concerning which of these aspects need improvement for particular audiences. This section finally collects overarching needs. 

 The need for general and uncertainty specific pedagogical knowledge was expressed in stakeholder letters. 
 Specifically NMIs
 \begin{itemize}
	\item would like to strengthen those teaching and those requiring an understanding of uncertainty particularly outside calibration laboratories and outside areas covered by EA-4/02 \cite{EA-4/02},  i.e.\ product testing in legal and regulatory contexts, uncertainties in chemical and biological analysis as well as multivariate uncertainties [SL19]. 
	\item expressed the need for a basis that facilitates knowledge transfer and the understanding of uncertainty, such as exchange of teaching methods and practical experiments [SL19].
	\item	emphasized the importance of the Monte Carlo Method, of correlation between quantities and of making sophisticated statistics better accessible to technical staff [SL14]. 
\end{itemize}

 The general need to strengthen those teaching and those requiring an understanding of uncertainty was expressed for all areas of chemical measurements, e.g.\ medicine, food safety, environmental and climate protection [SL1], as well as in civil engineering research to improve the use of metrology for, e.g. concrete dams, buildings, structures, geotechnics, hydraulics, materials, transportation [SL10], and for manufacturing in semiconductor industry and of measuring instruments to improve the quality of their products. In particular, exchange and training material on coverage and confidence intervals for non-metrologists, on uncertainty contributions that are unknown or originate from time or climatic drifts, the use of prior knowledge from similar products and examples for electrical high-frequency applications are of interest for the latter [SL24]. One NMI observed the need for a comprehensive, easy and efficient treatment of uncertainty in regression problems to support teaching in calibration contexts \cite{MUTPresINRIM}. (Also JCGM has been planning to publish a guide on applications of the least-squares method for some time \cite{JCGMNews}.) The European Commission's science and knowledge service aims to gain more visibility for their uncertainty training on reference materials and with that for the importance of reliable measurement results and the role of reference materials, e.g.\ in food, feed and environmental analysis, engineering and health applications [SL35]. A DI would like to strengthen its teaching on uncertainty due to significant retirement of teaching personnel. [SL22] In countries with fewer metrology resources and smaller audiences for uncertainty topics, it may be more difficult for NMIs and other training providers to offer uncertainty training that is tailored to heterogeneous audiences and contexts, and to do so regularly (see e.g.\ \cite{MUTPresIPQ,MUTPresSMD} for hints).

  Additional needs for uncertainty training may exist in communities that were not surveyed in this article.

\subsection{Emerging Needs}\label{subs:NeedEmerg}

 The EMN Mathmet identified in its strategic research agenda \cite{MathmetSRA} urgent needs, new challenges and opportunities in the areas of mathematics and statistics in metrology. For this purpose,
 Mathmet identified the strategic topics `Artificial Intelligence (AI) and Machine Learning (ML)' and `Computational Modelling and Virtual Metrology', as well as the foundational topic `Data Analysis and Uncertainty Evaluation' based on an extensive stakeholder consultation process. 
For the topic `Data Analysis and Uncertainty Evaluation', Mathmet foresees to face the following challenges relevant to measurement uncertainty: uncertainty evaluation for scientific applications, for small sample sizes and large sample sizes,	Bayesian statistics, analysis of key comparison data, statistical tests, as well as model design, selection and validation, accounting for model errors, and the GUM suite of documents. Persons interviewed by Mathmet reported, among others, the following applications for which guidance on appropriate uncertainty calculations was still missing: biomedical applications, high frequency applications (non-stable process), regression problems with complicated uncertainty structures, analytical measurements (chemistry and life sciences, non-stationary data), optical surface and coordinate metrology, as well as uncertainty calculations of AI and ML data analysis approaches. 

 For the topic `AI and ML' one of the recurring emerging issues relevant to uncertainty is its modelling and evaluation, in order to assure robustness and reliability, and, ultimately, the trustworthiness and traceability of results. Uncertainty quantification of AI/ML predictions is important for verification and validation of algorithms and software; understanding the impact of measurement uncertainty on robustness of AI/ML systems is essential to pave the way to their standardization and to improve decision making based on such systems; assessing confidence in AI and ML results is needed for certifying relevant applications. The presentation \cite{MUTPresNPL2} derives some training needs from current and emerging uncertainty approaches in ML.

 Also for the topic `Computational Modelling and Virtual Metrology' uncertainty evaluation is essential and there are still issues to be resolved in this context, especially when ensuring compliance with current standards in metrology such as the GUM \cite{JCGM100,JCGM101,JCGM102}.

 Most of the above-mentioned challenges and gaps, concerning the modelling and evaluation of measurement uncertainty, are not directly accompanied by an explicit need for a corresponding uncertainty training. Nonetheless, it is clear that any new challenge or research topic that is to be tackled, in a certain area or for a certain application, would need also some corresponding kind of support from the training side. Hence, ideas and strategies for the development of future training courses and curricula should be developed in connection with the plethora of present and emerging needs for measurement uncertainty evaluation, in the several application fields and for the several audiences that ask for support. 

\section{Conclusions}\label{sec:Concl}

 \subsection{Three Common Needs for Training on Uncertainty}
  This work revealed three major needs on uncertainty training that are common to many of the communities requiring an understanding of uncertainty.
	
  First of all, there is a need to address a general lack of training on uncertainty for calibration laboratories, testing laboratories, technical assessors of these, in legal metrology, in the field of the EMNs Climate and Ocean Observation, Advanced Manufacturing as well as Smart Electricity Grids, in the thermometry, length and analytical chemistry community, as well as for emerging topics. The lack is often identified for tailored approaches to uncertainty training (including examples), but also for fundamental or common approaches, and when new guidance or regulation is adopted. Smaller and emerging NMIs are in need of qualified teachers to strengthen their teaching or expertise. Calibration laboratories and teachers of uncertainty see the need to improve general and uncertainty specific pedagogical knowledge, e.g.\ by encouraging exchange between those teaching uncertainty. 
	
	Secondly, a better overview of the state of the art of uncertainty training is needed, especially at universities, for accreditation body personnel and in the different metrology fields represented by European Metrology Networks. Possibly also in legal metrology a better overview on uncertainty training is needed, if the lack of evidence on courses accessible to legal metrology audiences is not entirely due to an overall lack of such courses. Furthermore, all communities that need an understanding of uncertainty, but are not covered in this article, may benefit from an overview on the state of the art of uncertainty training.
	
	Thirdly, among the training that is provided, the information collected and investigated for this study highlights the need to deliver more training on specific technical topics related to evaluating uncertainty. In particular, little training on the propagation of distributions via Monte Carlo is provided for NMI staff, for calibration and testing laboratories and their assessors, as well as in legal metrology. This lack includes the validation of the law of propagation of uncertainty against Monte Carlo results. In addition, little training on multivariate measurement models is provided for NMI staff, for calibration and testing laboratories and their assessors as well as in legal metrology and at universities. Because correlation between input and/or output quantities is not rare, there is the need to raise awareness about the extension of propagating uncertainties and distributions to several output quantities in \cite{JCGM102} among these communities. In different communities, also training on other aspects of the evaluation of uncertainty is needed, e.g.\ on model building, uncertainty for sampling or evaluating input and combined uncertainties.		
 
 \subsection{Discussion and Outlook}
 
	To review the state of the art of, and the needs for, training on measurement uncertainty in Europe, this article merged many different sources of information. Some of these sources were raised within the activity MU Training between 2021 and 2024, and range from surveys, interviews and stakeholder letters to workshop presentations. Other sources range from research to websites on courses or strategic agendas. The authors do not claim complete or equal coverage of all training on measurement uncertainty in Europe, and caution about potential limitations.

  The needs identified for each community, as well as the three most common needs, indicate where future developments are required most to improve the understanding of uncertainty. 
	
	Some of these prioritized needs may be taken up and addressed by the MU Training activity that will continue in the future. This article may also guide future developments by EURAMET's Technical Committees and European Metrology Networks for their communities. For example, the EMN Mathmet is currently planning a summer school on uncertainty. In addition, training providers in all the communities covered and not covered here have the opportunity to take advantage of this study. 
	
	This article also provides a valuable source of information for teachers of uncertainty to better address the needs of their audience, to learn about research, practical teaching experience as well as available material and tools specific to uncertainty training. When implemented, this information will facilitate to improve the training on uncertainty.
	
	 Ultimately, this work will contribute to increasing the understanding of measurement uncertainty.
		
\section{Acknowledgements} 
 The authors would like to thank the consortium of the Measurement Uncertainty (MU) Training activity of the European Metrology Network for Mathematics and Statistics (Mathmet). This publication is an output of the activity, to which all partners contributed.

 In particular, the authors would like to thank ACCREDIA (in particular Paola Pedone, Sabrina Pepa and Fabrizio Manta) for designing and conducting the questionnaire described in \ref{AppQuest} and making the results available for this work. 

 We would also like to thank John Greenwood (UKAS) and Cord M\"uller (DAM) for providing informal feedback to draft versions of selected sections of the manuscript.

 In addition, we would like to thank Thierry Caebergs (SMD) and Massimo Ortolano (Politecnico di Torino) for providing input at an early stage of this study.

NPL’s work was supported by the UK Government’s Department for Science, Innovation and Technology (DSIT) as part of its National Measurement System (NMS) programme.\vspace{-1.6ex}

\section*{References}\vspace{-0.5ex}
 \bibliographystyle{iopart-num}
 \bibliography{MostCommonNeedsMUTraining}

\appendix
\section{Description of Sources}\label{AppSource}
 \subsection{Course survey}\label{AppSurvey}
  A review of courses on measurement uncertainty \cite{CourseSurvey}, usually offered by the partners of the consortium of the MU training activity and some of its stakeholders \cite{MUTrain}, was performed in order to i) provide the state of the art of measurement uncertainty training at a European level, ii) inform the wider audience about the characteristics of the available courses, and iii) identify possible gaps in the provided education. The courses were categorized according to their main features (language, technical level, audience, duration etc.) and, in particular, according to their adherence with the prescriptions of the GUM suite of documents \cite{JCGM100,JCGM101,JCGM102,JCGM106,GUM6}, with a special focus on the teaching of the Law of Propagation of Uncertainty (LPU, \cite{JCGM100,JCGM102}) and the Monte Carlo Method (MCM, \cite{JCGM101,JCGM102}) for the propagation of probability distributions. 
	
	An analysis of the first 41 collected courses is available at \cite{Penn2023}, thereafter the survey was extended by two more courses. Of all the 43 courses$^1$, 35 \% are aimed at NMI staff, 51 \% at calibration and testing laboratories, 14 \% at legal metrology staff, and 28 \% at academia. More detailed information about the proportions of courses aimed at different (non-overlapping) target audiences among the 43 courses collected in the survey is reported in table~\ref{FigTargetAudience}. 
	
		\begin{table*}
		\caption{\label{FigTargetAudience}Proportions of courses aimed at different target audiences among the 43 courses collected in the survey \cite{CourseSurvey}.}
			\begin{indented}\item[]
			\begin{tabular}{@{}c|c@{}}
			target audience for courses & proportion\\\hline
			NMIs & 14\%\\
			NMIs, calibration \& testing labs & 21\%\\
			calibration \& testing labs& 23\%\\
			academia, calibration \& testing labs & 5\%\\
			academia & 23\%\\
			legal metrology & 12\%\\
			legal metr., calibration \& testing labs & 2\%
			\end{tabular}
			\end{indented}
		\end{table*}

	The 2023 version of the course survey \cite{CourseSurvey} was used for this article, but will be updated regularly.
 	
	\subsection{Questionnaire for accredited laboratories and their assessors}\label{AppQuest}
	 A questionnaire inquiring about the interest, the (self-reported) knowledge and the preferred teaching method on 21 topics related to measurement uncertainty was designed by ACCREDIA (lead), a focus group and partners of the MU Training activity. The focus group included a sample of technical assessors and accredited laboratory personnel, both from testing and calibration. For the design of the questionnaire separate sessions were conducted for each group.	At the beginning of 2022 the questionnaire was distributed to members of the European co-operation for Accreditation, EA, and to accredited laboratories in Italy as well as their assessors. Table \ref{TabAccredPartic} lists the different numbers of participants, with a total number of 805 respondents for Italy and a response rate of 46 \% there, and 476 respondents for EA with a considerably lower rate.\footnote{\label{note2}Respondents working in both calibration and testing or in the laboratory and as assessor are listed multiple times in table~\ref{TabAccredPartic}.} 
		
		\begin{table*}
		\caption{\label{TabAccredPartic}Numbers of participants responding to a questionnaire on measurement uncertainty topics among different EA (and in parentheses Italian) accredited laboratories and their assessors.}
		\begin{indented}\item[]\vspace{0.5ex}
			\begin{tabular}{@{}l|ll@{}}
			 & calibration & testing \\\hline
			 accredited lab. & 82 (133) & 258 (547)\\
			 techn. assessor& 59 (38) & 140 (155)
			\end{tabular}
		 \end{indented}
		\end{table*}
		
	 The uncertainty topics about which the questionnaire inquired are listed in table~\ref{TabAccredTopics}. Question 19 on `Uncertainty factor, new approaches for expanded measurement uncertainty evaluation' was excluded from further analyses in sections~\ref{sec:stateofart} and~\ref{sec:needs} due to its ambiguity. Table~\ref{TabAccredTopics} also displays the average knowledge and interest scores that participants of different communities reported on these 21 topics.
	
	\subsection{WELMEC consultation}\label{AppWelm}
	The European Cooperation in Legal Metrology, WELMEC, had 9 active working groups (WGs) in 2022. Representatives of 8 of these WGs, i.e.\ WGs 2, 6-8 and 10-13, were interviewed on their understanding and their need for training on measurement uncertainty. This consultation was conducted by IMBiH (lead) and other partners of the MU Training activity. 
	
	For the consultation, the representatives of all WELMEC WGs were provided with the questions listed in table~\ref{TabWelmecQ}. One WG answered in writing, while the other 7 WGs shared their opinion during a subsequent meeting. The WG convenors are usually NMI employees and stated to have themselves appropriate knowledge on measurement uncertainty. The main topic of interest among the WGs is conformity assessment, where WG 8 `General Application of MID and NAWID' stated to welcome support for implementing JCGM 106 \cite{JCGM106} which accounts for uncertainties and risks. An additional discussion took place with representatives of WG 6 `Prepackages', who expressed that training on model building could be useful when changes to the measurement model for the verification of prepackages in WELMEC Guide 6.9 \cite{WELMEC6.9} are required, e.g.\ due to new regulations or technology. In addition, WG 6 is interested in  software for uncertainty evaluation.

	\begin{table*}
	 \caption{\label{TabWelmecQ}List of questions provided to WELMEC WGs prior to a meeting between its representatives and partners of the MU Training activity.}
	 \begin{indented}\item[]\hspace{-7em}
	 \begin{tabular}{@{}lp{0.95\textwidth}@{}}
	\\[-1.5ex]\hline
	1 & How is your involvement regarding measurement uncertainty best described?\\
	2 & Which literature/guides for uncertainty evaluation does your WG use most?\\
	3 & What are the biggest challenges you face when evaluating measurement uncertainty?\\
	4 & Are there applications for which guidance for appropriate uncertainty calculations is missing?\\
	5 & Would you be interested in measurement uncertainty training, please specify which ones?\\
	6 & What kind of research projects related to uncertainty evaluation would be useful for you?\\
	7 & Do you have any specific expectations from EMN MATHMET, or what could the EMN do to help you in your work with uncertainties?\\\hline
	 \end{tabular}
	 \end{indented}
	\end{table*}
		
	\subsection{Stakeholders and partners of the MU Training activity}\label{AppSL}
	There are 34 stakeholders who expressed interest in the activity MU Training. Their formal letters of interest are labelled [SLn] in this study and were received mostly in 2021 before the activity started; some were received during the activity and others as late as 2023. The stakeholders are listed on the activity website \cite{MUTStakeh} and range from one-person companies to large cooperations or networks to well-known industrial partners. Their views and needs differ widely and may or may not be representative for some community.	The views of one, non-European stakeholder are not included in this study. 
	
  The 16 organizations contributing to the MU Training activity are listed on the activity website \cite{MUTrain}. At the start of the activity a protocol was formulated, in which each partner described, among others, their ability, status and partially also need for teaching uncertainty. Some of these statements are included in this study and labelled [PS].
	
	\subsection{Survey of EMNs for different metrology fields}\label{AppEMN}
	For the metrology fields of Advanced Manufacturing, Clean Energy, Climate and Ocean Observation, Energy Gases, Laboratory Medicine, Mathematics and Statistics, Pollution Monitoring, Quantum Technologies, Radiation Protection, Safe and Sustainable Food, Smart Electricity Grids as well as Smart Specialisation in Northern Europe, documents and public information of the corresponding EMNs were analysed with respect to the key words `uncertainty' and `training' at the beginning of 2024. The strategic research agenda of 6 of the EMNs either does not mention these keywords or is not available yet. The needs regarding uncertainty and uncertainty training that the remaining EMNs (Advanced Manufacturing, Climate and Ocean Observation, Laboratory Medicine, Mathematics and Statistics, Radiation Protection and Smart Electricity Grids) identified are described in section~\ref{subs:EMNs}. These needs usually reflect priorities of the whole community and its stakeholders.

\end{document}